\documentclass[sigconf,nonacm]{acmart}
\acmSubmissionID{1025s1}

\usepackage{booktabs} 

\usepackage{url} 

\usepackage[ruled]{algorithm2e} 
\usepackage{wrapfig}

\SetAlFnt{\small}
\SetAlCapFnt{\small}
\SetAlCapNameFnt{\small}
\SetAlCapHSkip{0pt}

\acmJournal{TOG}

\begin{document}
\title{View2CAD: Reconstructing View-Centric CAD Models from Single RGB-D Scans}

\author{James Noeckel}
\orcid{0000-0002-6519-0322}
\affiliation{%
  \institution{University of Washington}
  \city{Seattle}
  \state{WA}
  \country{USA}}
\email{jamesn8@cs.washington.edu}
\author{Benjamin Jones}
\orcid{0000-0001-8524-4730}
\affiliation{%
	\institution{Massachusetts Institute of Technology}
	\city{Cambridge}
	\state{MA}
	\country{USA}}
\email{bt\_jones@mit.edu}
\author{Adriana Schulz}
\orcid{0000-0002-2464-0876}
\affiliation{%
	\institution{University of Washington}
	\city{Seattle}
	\state{WA}
	\country{USA}}
\email{adriana@cs.washington.edu}
\author{Brian Curless}
\orcid{0000-0002-0095-5400}
\affiliation{%
	\institution{University of Washington}
	\city{Seattle}
	\state{WA}
	\country{USA}}
\email{adriana@cs.washington.edu}



\begin{abstract}
Parametric CAD models, represented as Boundary Representations (B-reps), are foundational to modern design and manufacturing workflows, offering the precision and topological breakdown required for downstream tasks such as analysis, editing, and fabrication. However, B-Reps are often inaccessible due to conversion to more standardized, less expressive geometry formats. Existing methods to recover B-Reps from measured data require complete, noise-free 3D data, which are laborious to obtain. We alleviate this difficulty by enabling the precise reconstruction of CAD shapes from a single RGB-D image. We propose a method that addresses the challenge of reconstructing only the observed geometry from a single view. To allow for these partial observations, and to avoid hallucinating incorrect geometry, we introduce a novel view-centric B-rep (VB-Rep) representation, which incorporates structures to handle visibility limits and encode geometric uncertainty. We combine panoptic image segmentation with iterative geometric optimization to refine and improve the reconstruction process. Our results demonstrate high-quality reconstruction on synthetic and real RGB-D data, showing that our method can bridge the reality gap.
\end{abstract}

%
%

\ccsdesc[500]{Computing methodologies~Shape modeling}

%
%

\keywords{3D Reconstruction, Computer-Aided-Design, Reverse Engineering}

\newcommand{\rgbimage}{\mathcal{I}_{\mathrm{rgb}}}
\newcommand{\depthimage}{\mathcal{I}_{\mathrm{D}}}
\newcommand{\truevbrep}{\mathcal{V}}
\newcommand{\truebrep}{\mathcal{B}}
\newcommand{\pointcloud}{\mathbf{P}_{\mathrm{depth}}}
\newcommand{\points}{P}
\newcommand{\cleanpoints}{\tilde{P}}
\newcommand{\surface}{\mathbf{S}}
\newcommand{\visiblesurfaces}{\mathcal{S}}

\newcommand{\silhouette}{silhouette}
\newcommand{\visibility}{visibility}
\newcommand{\intersection}{intersection}
\newcommand{\occluded}{occluded}

\newcommand{\Silhouette}{Silhouette}
\newcommand{\Visibility}{Visibility}
\newcommand{\Intersection}{Intersection}
\newcommand{\Occluded}{Occluded}

\newcommand{\intthresh}{d_{\mathrm{int}}}
\newcommand{\ransacthresh}{d_{\mathrm{inlier}}}
\newcommand{\edgemap}{\mathcal{E}}
\newcommand{\edgemaplifted}{\mathcal{E}_{\mathrm{3D}}}
\newcommand{\edgepoints}{\mathbf{V}_w}
\newcommand{\point}{\mathbf{v}}
\newcommand{\pointtd}{\mathbf{q}}
\newcommand{\graphedge}{\mathbf{E}_w}
\newcommand{\initialedgepoints}{\mathbf{V}_0}
\newcommand{\liftededgepoints}{\mathbf{V}_{\mathrm{3D}}}
\newcommand{\intersectionmask}{I}
\newcommand{\visiblemesh}{\mathcal{M}}
\newcommand{\neighbors}{\hat{N}}

\newif\ifshowauthorcomments
\showauthorcommentsfalse
\newcommand{\authorcomment}[3]{\ifshowauthorcomments{\bfseries \scriptsize \color{#3} #1: #2}\fi}
\newcommand{\adriana}[1]{\authorcomment{AS}{#1}{red}}
\newcommand{\ben}[1]{\authorcomment{BJ}{#1}{magenta}}
\newcommand{\james}[1]{\authorcomment{JN}{#1}{green}}
\newcommand{\brian}[1]{\authorcomment{BC}{#1}{blue}}

\begin{teaserfigure}
    \centering
    \includegraphics[width=\linewidth]{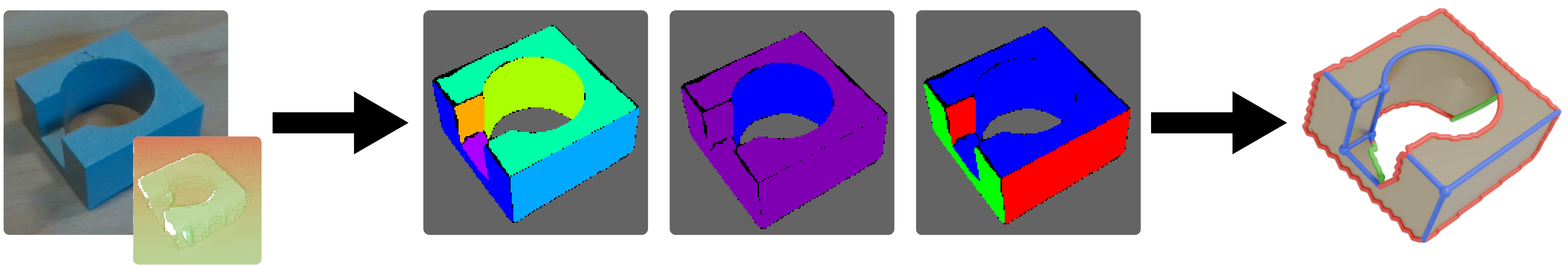}
    \caption{We extract view-centric CAD geometry from a single noisy real-world RGB-D scan. We achieve high precision on partially observed data by learning image-space geometric priors (middle) to aid in constrained optimization for the final view-space geometry. Our view-centric output geometry (right) identifies edges/vertices that belong to the original CAD model with high certainty (blue) and distinguishes them from silhouette edges or occlusion effects (red and green). }
    \label{fig:teaser}
\end{teaserfigure}

\maketitle

\section{Introduction}



Computer-Aided Design (CAD) tools are foundational to the creation of nearly all man-made objects, allowing designers to precisely specify models that are directly converted into manufacturing plans. These CAD models are typically stored as Boundary Representations (B-Reps), a format that provides high precision and facilitates further editing---critical features for design and manufacturing workflows.

Despite their prevalence, B-Rep models are often inaccessible due to proprietary restrictions or conversion into less precise formats like meshes, which lose fidelity and editability. Additionally, industrial objects are frequently modified during manufacturing to address unforeseen requirements, leaving original CAD models outdated. These challenges create an urgent need for reconstructing B-reps from existing objects, enabling them to be seamlessly integrated into modern design pipelines. This would support analysis, quality control, and the creation of new designs built on precise, editable representations.

Reverse-engineering CAD models has been extensively studied, but reconstructing high-precision B-Reps remains a significant challenge. It requires not only capturing geometry but also understanding the composition of their edges, faces, and vertices. Most promising approaches rely on full, precise point clouds~\cite{buonamici_reverse_2018, liu2024point2cad, liu_sig24,sed_net_2023,sharma2020parsenet,li2019supervised}, which are often impractical to capture. In this work, we ask: can we reconstruct a high-precision B-rep from a single depth image? This opens possibilities for applications where complete point clouds are unavailable, making reconstruction more practical and widely applicable. For example, CAD enables designs to interface accurately with other objects by referencing B-rep topological entities~\cite{jones2023brepmatching}. By enabling in-situ capture of B-reps of existing objects, our method supporting tasks like designing and fabricating fixtures, attachments, or custom components that seamlessly integrate with them.  

\begin{wrapfigure}{r}{0.2\textwidth}
    \centering
    \vspace{-10pt}
    \includegraphics[width=0.2\textwidth]{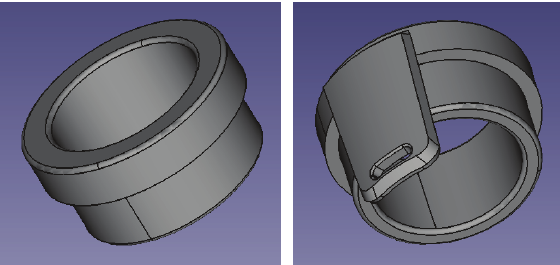}
    \vspace{-20pt}
    \label{fig:illustrative}
\end{wrapfigure} CAD models used in engineering often include features that are not immediately visible (see inline image). Because CAD geometry for mechanical parts needs high accuracy, usually to interact with some other part, hallucinated geometry is unlikely to be helpful and may even be harmful. Instead, we focus on capturing observed geometry while clearly distinguishing between high-confidence features and areas of uncertainty (see Figure~\ref{fig:teaser}).

A core challenge in reconstructing B-Reps from a single view is that the geometric priors required for B-Rep reconstruction rely on complete topological information. B-Reps represent geometry as a graph of faces, edges, and vertices, each defined by unbounded parametric functions. Accurate geometric interpretation requires the full topological graph, with vertices bounding curves to form edges and edges bounding surfaces to form faces. While prior methods leverage B-Rep priors with complete data, single-view input provides only partial and often ambiguous information; for instance, observed edges may represent true B-Rep edges or be effects of occlusion or limited visibility, complicating reconstruction.

Our work addresses these challenges with two key insights. The first is that local information independent of topology can be directly extracted from images. We propose to recover details about visible faces, including surface types and orientations, by framing the task as a \textit{panoptic segmentation} problem, which enables precise identification of regions associated with visible faces in the image. \adriana{ add that images can all tell us about global orientation }

The second insight is to model only the geometry we observe. We define a \textit{view-centric B-Rep} (\textbf{VB-Rep}), a modification of the standard CAD B-Rep, which introduces additional structures to handle visibility limits while preserving B-rep precision. These visibility boundaries encode geometric uncertainty from a single view, allowing us to differentiate between elements of the CAD model and artifacts from occlusions or silhouettes.  

We build on the VB-Rep’s representation of uncertainty to iteratively refine the reconstruction, focusing on high-confidence elements. By isolating edges likely originating from the CAD B-Rep and distinguishing them from visibility artifacts, we iteratively align and refine surface fits. This process yields additional edge information, enabling robust, feedback-driven improvements on incomplete data.

We demonstrate the effectiveness of our method on a collection of real objects captured with a commodity depth sensor, as well as a synthetic dataset of photorealistic RGB-D images. Through ablation studies, we highlight the importance of the various types of geometric guidance we extract through vision and optimization.

\section{Related Work}


Research related to our task can be arranged into three main categories: single-view reconstruction, CAD reverse engineering, and CAD generative modeling. 

\noindent \textbf{Single-View Reconstruction}
Single-view reconstruction is a problem that has been widely studied and recently has made great progress with advances on 3D generative models that can be conditioned on images~\cite{Park_2019_CVPR,choy_r2n2_2016,groueix2018}, effectively generating 3D representations from a single view. State of the art results are achieved with recent foundation models ~\cite{liu2023zero1to3, liu2024one,liu2024onep,xiang2024structured,bala2024edify}. Despite their ability to produce visually impressive results, the outputs these models generate, dense meshes or implicit neural representations, lack the precision and meaningful topological decomposition of B-reps, which break objects into well-defined faces, edges, and vertices. This level of detail is essential for tasks that demand accuracy and reliability, including downstream editing, manipulation, and fabrication.



\noindent \textbf{CAD Reconstruction}
Given the useful properties of CAD representations, many recent works have sought to reconstruct these directly from measured data. These works assume that the entire geometry of the object is observed in the form of a point cloud, and frame the problem as a segmentation task~\cite{li2019supervised,yan2021hpnet,sharma2020parsenet,zong2023p2cadnet,liu2024point2cad}. For B-Rep extraction, segment-and-fit methods like \cite{sed_net_2023} use intersection curves to infer topological edge structures, but their effectiveness depends on the quality of input data. Complexgen \cite{guo_complexgen_2022} jointly predicts geometry and topology but still requires optimization to fit parametric surfaces. Other methods, such as Split-And-Fit \cite{liu_sig24}, are effective only on synthetic, uniformly sampled point clouds, limiting applicability to real-world, non-uniform data. Another way of representing CAD models is using construction sequences such as constructive solid geometry. Methods for reconstructing in this space also require full mesh/point data as input~\cite{du2018inversecsg}. Clean 3D point clouds or meshes needed as input to these methods can be expensive to obtain with sufficient detail for reconstruction. In this work, we enable CAD reconstruction with just a single view and approximate depth information. 

While some robotics-focused works have explored incremental B-Rep reconstruction from single-view depth images~\cite{sand2016brep, sand2017matching}, these only handle simple planar geometries and rely strongly on high-quality depth images, as it is the only input. Our approach leverages deep learning with RGB data to infer priors that enable reconstruction of detailed features not captured in the depth map.

\noindent \textbf{Generative CAD}
Another related direction of research is in generative modeling of CAD representations. These models can be trained with image conditioning in order to perform reconstruction. Many CAD-focused generative models tend to be limited to the simple geometric domain of sketch-extrude CAD~\cite{wu2021deepcad,xu2022skexgen,you2024img2cad, chen2024img2cad,khan2024cad}, or otherwise restricted to a learned latent space~\cite{xu2024brepgen,jayaraman2023solidgen}. Additionally, these models do not condition well on image input, meaning that input images serve as weak guidance in practice. Concurrent work \cite{chen2024img2cad,alam2024gencad,xu2024cad} generates CAD geometry that closely follows input images,  but this has only been demonstrated for very simple geometry, and the input images must be identically styled schematic renders, limiting the generality of the method.

To our knowledge, this work is the first to reverse engineer B-Reps from single-view RGB-D inputs, and do so from low-precision depth sensors.

\section{View-centric B-Reps}\label{sec:brep}
\begin{figure}[h!]
    \centering
    \includegraphics[width=0.75\linewidth]{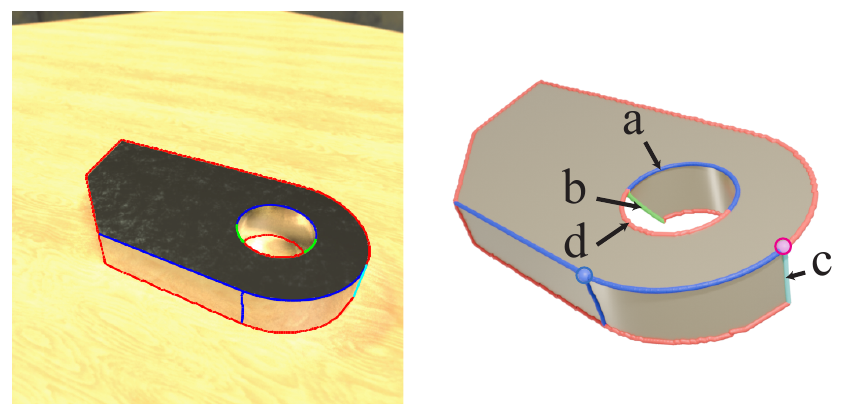}
    \caption{
    The visible surfaces subtend regions in the 2D image whose boundaries form a planar graph (left). The corresponding VB-Rep edges (right) are classified as \intersection\ (a), \occluded\ (b), and \visibility\ (c). All others we label as \silhouette\ edges (d). The blue vertex is a triple intersection, and the magenta vertex is an \intersection-\visibility\ vertex.}
    \label{fig:vertices}
\end{figure}

The subset of a B-Rep visible from a single view lacks topological validity—for instance, we might detect a surface primitive without its bounding curves. However, in view-space, the boundaries between the visible surfaces form a planar graph of edges that partitions 2D space into bounded regions. Some of these edges correspond to true B-rep edges, while others are visibility artifacts, but all define the visible limits of surfaces (see Figure~\ref{fig:vertices}).

\begin{figure*}[ht!]
    \centering
    \includegraphics[width=0.8\textwidth]{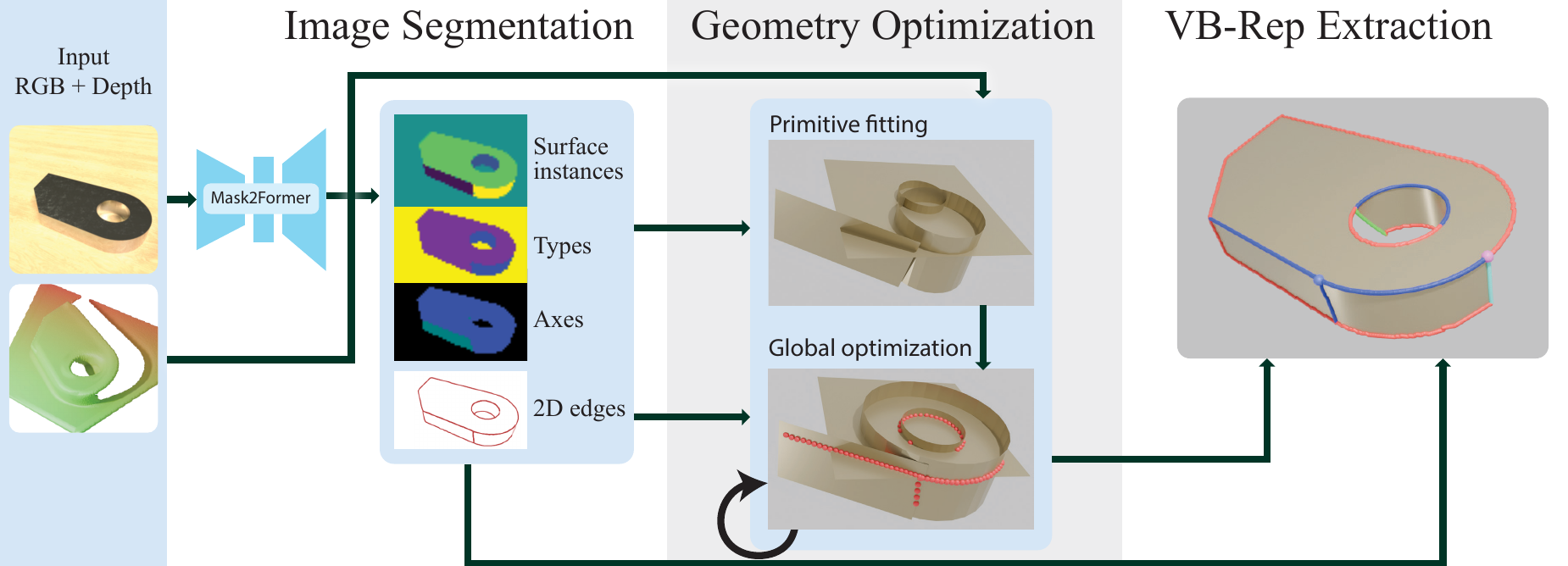}
    \caption{An overview of our pipeline. In the \textbf{segmentation stage}, we train a Mask2Former model~\cite{cheng2021mask2former} to perform panoptic segmentation of the RGB image into primitive instances along with classifying them according to primitive type and axis alignment. In the \textbf{geometry optimization} stage, we use the predicted instance masks and types to fit primitives to the input depth points, and optimize them to enforce consistency with predicted orientation and intersection constraints. Furthermore, we \textit{iterate} this step to update our knowledge of which edges to use as guidance in the optimization. Finally, in the \textbf{VB-Rep extraction stage}, we use the refined surface primitives and edges to build a coherent CAD representation consisting of bounded surfaces, curves, and points, labeled according to visibility information.}
    \label{fig:pipeline}
\end{figure*}

Motivated by this structure, we introduce the \textit{View-centric B-Rep (VB-Rep)}, which integrates visibility into the boundary graph of a standard B-rep. Similar to a B-rep, a VB-Rep is a graph of faces, edges, and vertices. However, edges of VB-reps are categorized by boundary source: \textbf{\intersection}, \textbf{\visibility}, \textbf{\occluded}, and \textbf{\silhouette} edges. \Intersection\ edges are true B-rep edges, defined by the intersection of two visible B-rep faces. \Visibility\ edges mark visible limits of primitives, while \occluded\ edges occur when there is a nearer occluding surface, and a corresponding occluding edge (edges (b) and (d) in Figure \ref{fig:vertices}). \Silhouette\ edges are all others not covered by these three. 

VB-Reps have two main advantages: they enable visualization of partially reconstructed CAD models by preserving valid bounds, and they allow assessment of which boundary elements (edges and vertices) likely belong to the true B-Rep. This helps guide CAD reconstruction by identifying where surfaces should intersect based on observed edges.

\section{Method}
\label{sec:methods}
Our method reconstructs a VB-Rep $\truevbrep$ from single-view RGB-D data $(\rgbimage, \depthimage)$ of a CAD object represented by a true B-Rep $\truebrep$. The depth capture parameters are known, so we obtain a frontal point cloud $\pointcloud$. Shown in Figure \ref{fig:pipeline}, our pipeline has three stages: First, in the \textbf{segmentation} stage, we perform panoptic segmentation of $\rgbimage$, identifying visible surface instances $\visiblesurfaces$ along with their types and axis alignments. Next, in \textbf{geometry optimization}, we fit primitives to the point cloud $\pointcloud$ derived from $\depthimage$, iteratively refining their orientation and enforcing intersection constraints to approximate $\truevbrep$. Finally, we construct a VB-Rep by extracting a wireframe representation, assigning edge types like \occluded, \visibility, and \intersection, thereby creating a coherent CAD model suitable for partial reconstruction from real-world RGB-D captures.

\subsection{Primitive Segmentation}

To reconstruct VB-Reps with finer detail than the noisy depth map alone can capture, we leverage high-quality RGB data to detect visible surface primitives. By framing primitive detection as a \textbf{panoptic segmentation} task on $\rgbimage$, we use a deep segmentation model to simultaneously identify all distinct visible parametric surfaces $\visiblesurfaces$ and classify each. More formally, given the image $\rgbimage$, we obtain a set of \textit{masks} $\mathbf M_i,\ i\in[1\ldots K]$ corresponding to $K$ detected surface instances, along with a \textit{type vector} $\mathbf{T_i} \in {0\ldots N_{\mathrm{types}}}$ specifying the type of each instance.

\begin{figure}[h!]
    \centering
    \includegraphics[width=0.5\textwidth]{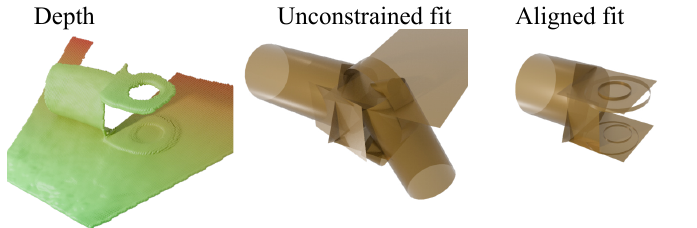}
    \caption{Fitting primitives to a noisy point cloud. Fitting primitives individually (middle) fails to recover the correct orientations especially in the thin cylindrical borders where point supervision is sparse. Meanwhile, globally aligned fitting (right) allows the thin structures to be correctly oriented with the same level of noise in the points.}
    \label{fig:noisy}
\end{figure}

For determining the classification types in our segmentation task, one approach would be to follow \cite{li2019supervised} and similar methods by predicting the types of each surface primitive for eventual fitting to $\pointcloud$; however, such fits are highly sensitive to noise, especially for small features like fillets, extrusions, or surfaces viewed from steep angles (see Figure~\ref{fig:noisy}). We address this by observing that a significant number of features in CAD B-Reps (and man-made objects in general) are aligned with coordinate axes, which serves as a strong geometric prior under uncertain depth estimates, so we include alignment in the learning objective. Deducing the orientation of a surface's underlying primitive axis is interdependent with identifying its type, so we treat axis alignment and surface type prediction as a \textit{joint} problem by defining each type label $\mathbf T$ to denote a primitive type combined with one of four orientation classes \textbf{(X, Y, Z, or unaligned)}.  We consider five primitive types---plane, cylinder, sphere, torus, and cone---with spheres having only one orientation, and a separate label for background, resulting in $18$ possible labels.

\begin{figure*}
    \centering
    \includegraphics[width=0.75\textwidth]{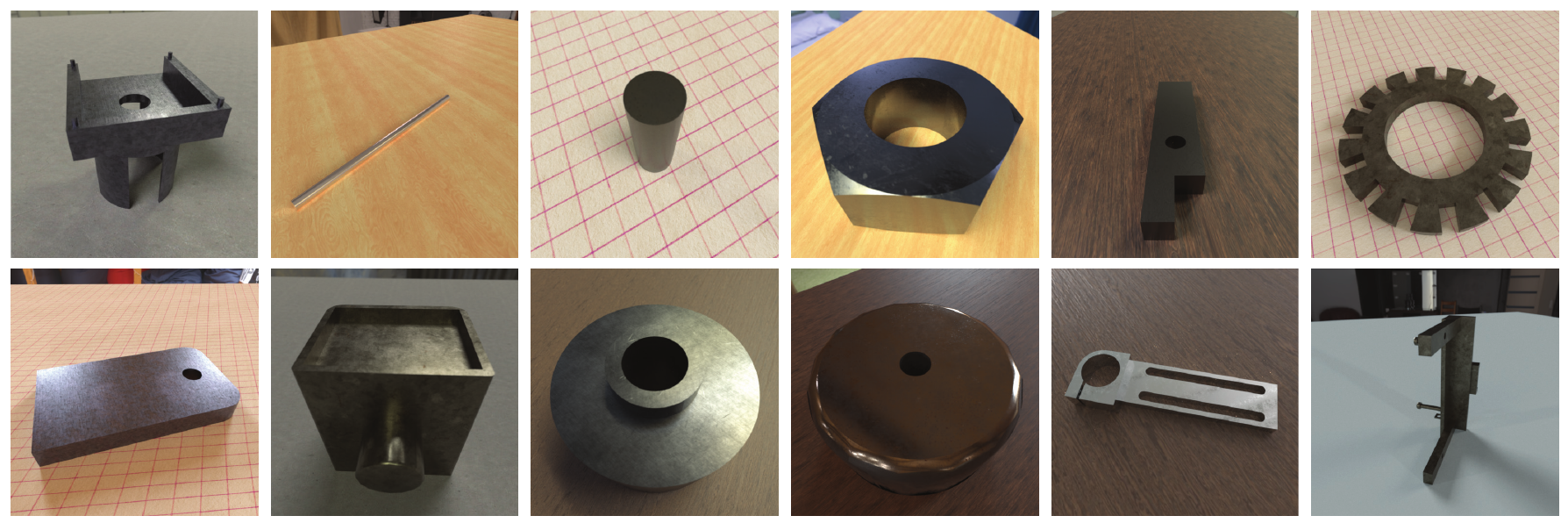}
    \caption{Representative images from synthetic training dataset.}
    \label{fig:dataset}
\end{figure*}

\vspace{2pt}\noindent\textbf{Dataset} We built a dataset comprising 50,000 synthetic renders of CAD models from the Onshape public repository~\cite{automate} featuring plane, cylinder, sphere, cone, and torus primitives with randomized lighting, materials, and camera viewpoints (see Figure~\ref{fig:dataset} for examples). Each render includes ground truth labels for primitive types and axis alignments, adjusted per view. Further details on scene setup and labeling are provided in Section~\ref*{sup:dataset} of the supplemental material.

\vspace{2pt}\noindent\textbf{Segmentation Network Implementation} We use the \path{maskformer2_R50_bs16_50ep} variant of the Mask2Former~\cite{cheng2021mask2former} model. The inclusion of an explicit background label helps to ensure accurate mask borders at silhouette boundaries.



\begin{figure} [h!]
    \centering
    \includegraphics[width=0.5\linewidth]{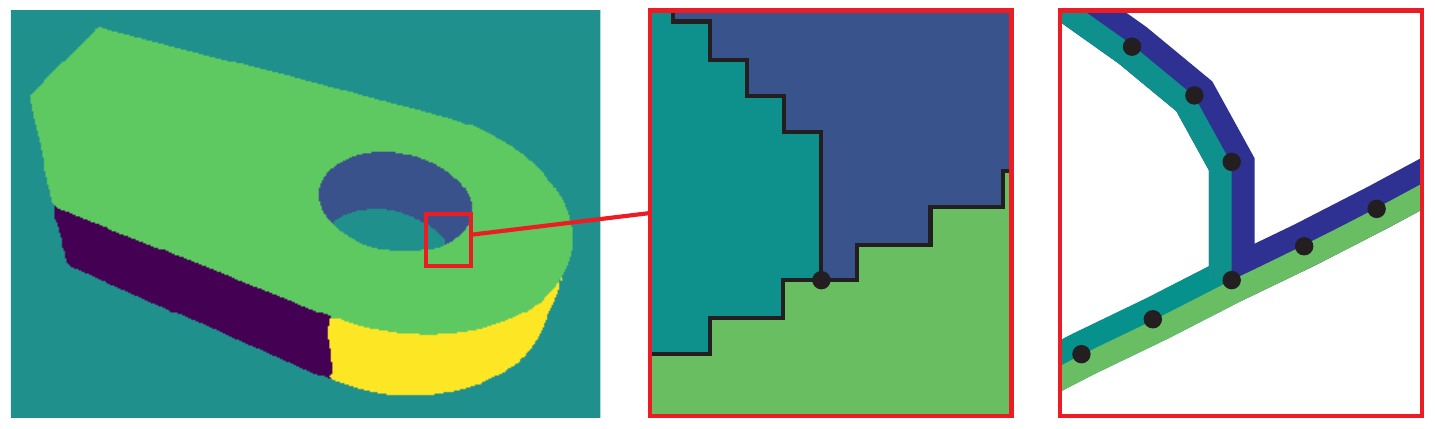}
    \caption{Given the initial instance map, we extract pixel-wise contours, then infer a simplified edge graph with adjacency information shown as dual edge colorings (right).}
    \label{fig:contours}
\end{figure}

\vspace{2pt}\noindent\textbf{2D Edge Graph Extraction} Given the segmentation, we extract a 2D edge graph from the boundary contours on the pixel grid between different instance labels and denote it $\edgemap=(\edgepoints, \graphedge)$, where the edges $\graphedge$ are line segments between the 2D vertices $\edgepoints$ (see Figure \ref{fig:contours}). We associate each edge $E_i \in \graphedge$ with its two neighboring instances 
in the segmentation map, indicated by the dual edge colors in Figure~\ref{fig:contours}. We define the set of neighboring surfaces for each vertex as all surfaces neighboring the adjacent edges and denote it by $\neighbors_i$ for $\point_i \in \edgepoints$.



\subsection{Geometry Optimization} 
\label{sec:geomopt}

Given the segmented labels and point cloud, our next step is to optimize the parameters of the surface primitives to align them with the predicted axis labels while fitting to the point cloud. To ensure topological consistency, we must not only optimize each surface independently but also ensure that they intersect along the \intersection\ edges of the VB-Rep $\truevbrep$. At this stage, however, we have not yet recovered these \intersection\ edges, as they become apparent only once the 3D geometry of the surface primitives is understood. To address this, we use an iterative approach: initializing geometry with local surface information, refining intersection constraints, and re-optimizing until convergence.

\vspace{2pt}\noindent\textbf{Constrained Primitive Fitting}
Our first step is to fit parameters for each surface primitive $\surface_i \in \visiblesurfaces$ to align with the labeled axis and match the depth image. Each primitive has an associated segmentation mask $\mathbf{M}_i$, which links it to a subset of points $\points_i \in \pointcloud$. While we have axis alignment labels, the exact orientation of the axis-aligned coordinate system is unknown and will be optimized jointly with the geometry in this step.

We initialize this step by fitting each primitive in isolation to match its associated point cloud. Following the RANSAC approach outlined in \cite{schnabel2007efficient}, we sample primitive parameters using random minimal subsets of points in $\points_i$. We retain the parameters with the highest number of inliers among $\points_i$ within $\ransacthresh$ of $\surface_i$ and discard outliers, resulting in a cleaned point set $\cleanpoints_i$. Some parameter estimates depend on point normals, which we estimate from the gradient of the depth map using a Sobel filter.

We use the initial primitive fits to globally optimize all primitives with an alignment axis to fit the point cloud subject to the resulting constraints. Since the 3D
axes themselves are unknown, we additionally optimize the
orientation of an orthogonal coordinate frame in this global optimization step. Because the constraints require aligning groups of primitive axes with one of the three orthogonal frame axes, we represent the orientation of the alignment coordinate frame as a single axis-angle rotation $\mathcal{R}$ with three degrees of freedom. We re-parameterize all aligned axes as functions of $\mathcal{R}$.

The alignment optimization objective function is therefore
\begin{equation}
    E_{\mathrm{recon}}(\mathbf \Theta, \mathcal{R}) = \sum_i\sum_{p\in\cleanpoints_i} \mathrm{dist}\left(\surface_i(\mathbf \Theta, \mathcal{R}), p\right)\label{eq:erecon}
\end{equation}
where $\mathbf \Theta$ are the non-constrained primitive parameters. 

\begin{figure}
    \centering
    \includegraphics[width=0.5\textwidth]{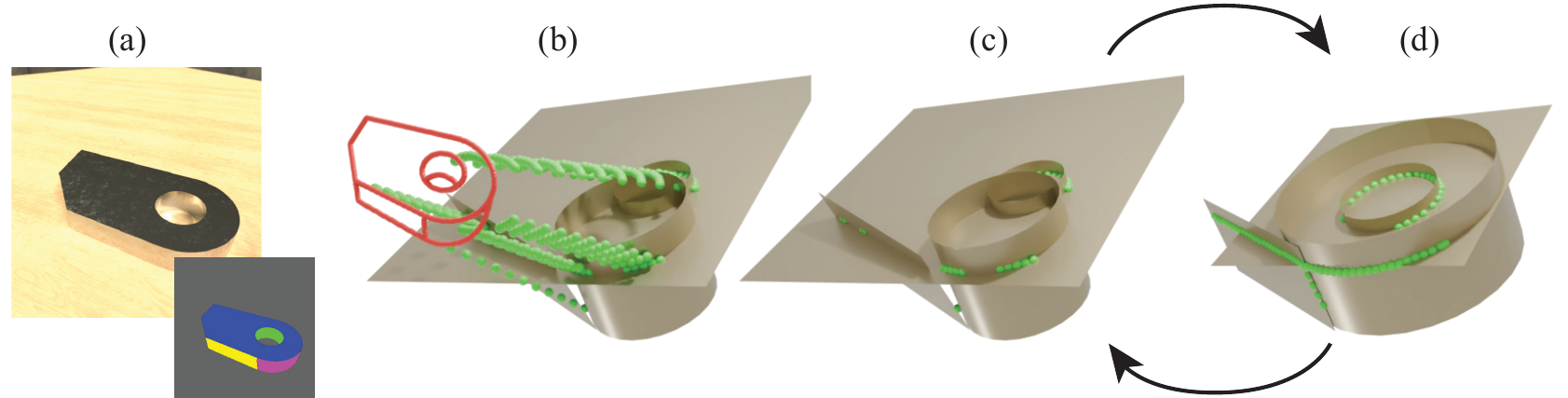}
    \caption{Overview of our intersection guidance optimization. Given the edges extracted from the segmentation map (a), we can minimize 
    the distance of the 3D vertices $s_i\hat\point_i$ to their neighboring surface primitives along the projected view direction (b), with $\hat\point_i$ shown in red. Edge points whose minimal distance is below $\intthresh$ are added to the intersection objective (c), shown in green. Finally, the surfaces and edge points are jointly optimized to minimize intersection distance while preserving the view projections of the edges (d); we also detect more intersection edge candidates (green) at this stage, so that we can repeat steps (c) and (d).}
    \label{fig:projection}
\end{figure}

\vspace{2pt}\noindent\textbf{Iterative Intersection Refinement}
Our next step is to refine the fit so primitives intersect along the edges in the 2D edge graph $\edgemap$ that correspond to intersections. Since $\edgemap$ exists in image space, it only specifies where surface intersections project onto the 2D plane. To guide primitive optimization in 3D, we define a corresponding 3D wireframe $\liftededgepoints(\mathbf{s}) = \{s_i\hat\point_i, i = 1 \dots |\edgepoints|\}$, where scalar variables $s_i$ control depth, and initial vertices $\hat\point_i$ are projected onto the 3D image plane using known camera intrinsics (see Figure~\ref{fig:projection} (b)). We can then define $E_{\mathrm{int}}$ as the total distance between each vertex $\pointtd_i\in\liftededgepoints$ and all its neighboring surfaces, $\neighbors_i$:
\begin{equation}
    E_{\mathrm{int}}(\mathbf{\Theta}, \mathbf{s}, \mathcal{R}) = \sum_i \intersectionmask_i\sum_k^{|\neighbors_i|}\mathrm{dist}(s_i\hat\point_i, \surface_{\neighbors_i^k}(\mathbf \Theta, \mathcal{R}))^2 \label{eq:eint}
\end{equation}
where $\intersectionmask_i\in\{0,\ 1\}$ is a label for whether each vertex lies at an intersection, indicating the subset of $\edgemap$ to consider, and $\mathcal{R}$, as before, controls the axis-aligned frame. In practice, we minimize $E_{\mathrm{int}} + w_{\mathrm{r}}E_{\mathrm{recon}}$ with $w_{\mathrm{r}}=0.1$ in this stage to avoid overfitting to edge artifacts.

Since the intersection edges are initially unknown, we use an iterative approach to alternately select $\intersectionmask_i$ and optimize primitive parameters. We first initialize the 3D positions of the wireframe vertices by  minimizing Eq.~\ref{eq:eint} with respect to only $s_i$, assuming $\intersectionmask$ is set to $\mathbf{1}$. For vertices where the final distance to neighboring primitives is below $\intthresh$, we set $\intersectionmask_i$ to 1, otherwise 0. Then, we minimize Eq.~\ref{eq:eint} with respect to both wireframe depths $\mathbf{s}$ and primitive parameters $\mathbf{\Theta}$ and then update $\intersectionmask$. This process is performed twice to capture intersection points initially missed (compare the points in Figure~\ref{fig:projection} (c) and (d)); we initially use threshold $\intthresh$ to discover as many intersections as possible with the unrefined surfaces, followed by $\intthresh/5$ in the second iteration to enhance accuracy.

In all optimization steps, we use the Levenberg-Marquardt algorithm with a dynamic damping factor as outlined in \cite{shakarji1998least}. We use parameter values $\intthresh=0.05$ and $\ransacthresh=0.03$, with our models scaled to have a maximum sidelength of 1.

\subsection{VB-Rep extraction}
\label{sec:wireframe}

\begin{figure*}
    \centering
    \includegraphics[width=0.9\textwidth]{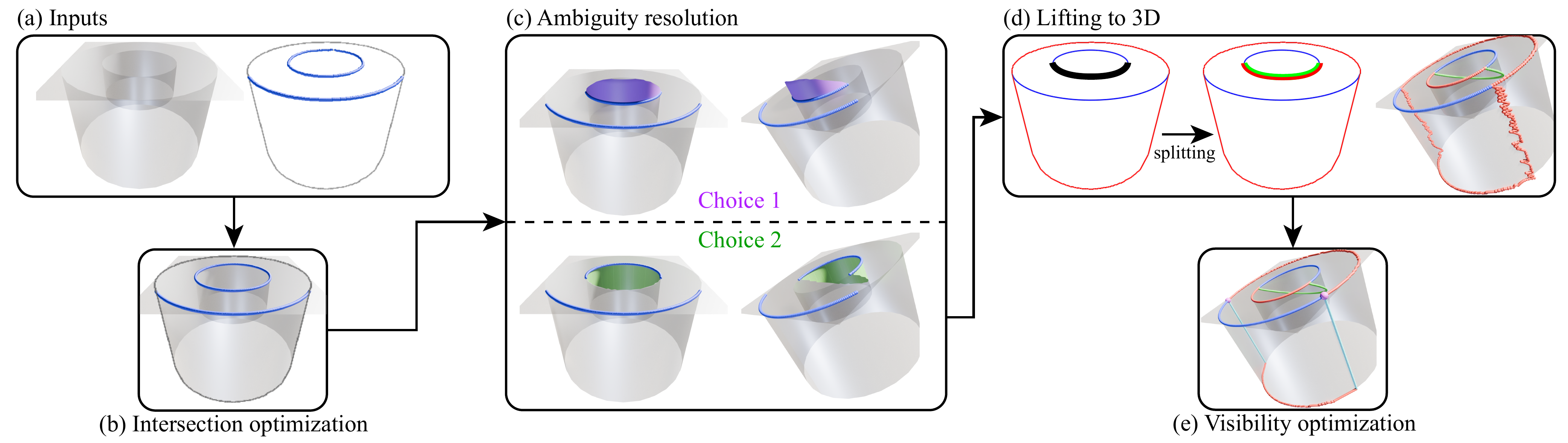}
    \caption{Surface-aware wireframe extraction pipeline. Input surfaces, 2D edge graph, and lifted intersection vertices (a) are optimized to the exact surface intersections (b), recovering potential missed intersections. We resolve ambiguities in how to lift the 2D boundaries into 3D (c) by comparing the resulting bounded surfaces, and keep the subset of intersection edges that agrees with that configuration. We can now split the wireframe along all non-intersection edges between surfaces (highlighed black edge in (d)), creating \silhouette\ edges (red) and \occluded\ edges (green), which we project to each surface to form a complete wireframe. Finally, we optimize the remaining \silhouette\ vertices to discover any visibility boundaries (e), shown in cyan.}
    \label{fig:wireframe}
\end{figure*}

After the previous steps, we have optimized surface primitives $\visiblesurfaces$ which will comprise the visible faces of our VB-Rep $\truevbrep$. To complete the structure, we also require edges to serve as boundaries for these faces, comprising a 3D wireframe. By the definition of a VB-Rep, these edges should all correspond to the observed 2D boundaries $\edgemap$. The main challenge of VB-Rep extraction is therefore correctly \textit{lifting} $\edgemap$ to 3D. We outline our procedure in Figure \ref{fig:wireframe}.


\vspace{2pt}\noindent\textbf{Intersection Optimization} The first step is to refine vertex positions in $\liftededgepoints$ by locally optimizing each point to minimize its distance to neighboring surfaces, allowing its full 3D position to vary freely. While this could be applied only to vertices marked as intersections in the previous step ($\intersectionmask_i = 1$), some vertices may be missed. 
Therefore, we perform this optimization for all points and compile a set $\mathcal{O}$ of vertices whose resulting positions are within $\intthresh$ of their neighboring 3D surfaces, provided they are within 5 pixels of their original projected image coordinates (see Figure \ref{fig:wireframe}~(b)). We note that not all points in $\mathcal{O}$ actually correspond to intersection edges in $\truevbrep$ because the \textit{visible} portions of the surfaces may not intersect along those points (see below).


\vspace{2pt}\noindent\textbf{Ambiguity resolution} For primitive surfaces that can have more than one depth from a given viewpoint, the interpretation of our boundaries so far may be ambiguous. This problem is illustrated in Figure \ref{fig:wireframe} (c), where the cylindrical surface may be interpreted as a protrusion (purple) or a hole (green). There are up to four different bounded subsets of a primitive surface with the same boundary when projected onto view space, which amounts to making a discrete choice. We construct view-centric meshes $\visiblemesh$ representing each of these choices and select the configuration whose mesh agrees most closely with $\pointcloud$. The procedure we use for extracting $\visiblemesh$ is detailed in Section \ref*{sec:mesh} of the supplemental material.

This process also allows us to determine which points in $\mathcal{O}$ are actually intersection edges in $\truevbrep$. As shown in Figure \ref{fig:wireframe} (c), only half of the edges in the inner circle correspond to a \textit{visible} intersection between the two primitives.


\vspace{2pt}\noindent\textbf{Lifting to 3D} 
After the disambiguation step, we identify which vertices in $\edgemap$ correspond to intersections. All other vertices either border a single surface or represent occlusions. For occlusions, we observe that two edges of $\truevbrep$---an \occluded\ edge and a \silhouette\ edge---will overlap in the 2D view. Since their projection will map to a single edge in $\edgemap$, we first split each vertex in $\edgemap$ identified as an occlusion into two separate points, assigning each to one of the two neighboring surfaces. We then update the edges accordingly.

With each remaining vertex now associated with a single surface, we project it onto that surface to complete the 3D boundaries, optimizing depths to minimize the distance to the surface. For surfaces with overlapping depths, we start by minimizing the distance to $\visiblemesh$ for disambiguation. Finally, we mark all newly split vertices as \textbf{\occluded} if their depth is greater than that of any overlapping vertices. The result of this process can be seen in Figure \ref{fig:wireframe} (d).

\vspace{2pt}\noindent\textbf{Visibility optimization} One last type of boundary remains: silhouettes which are solely caused by the intrinsic geometry of the surface, such as where a sphere or cylinder curves away from the view (Figure \ref{fig:vertices} (d)). Since the viewing direction is tangent to the surface along this boundary, the projected noise in the predicted silhouette tends to be amplified after the previous projection to 3D, causing a jagged boundary (Figure \ref{fig:wireframe} (d)). We make the observation, however, that there exists an exact mathematical visibility boundary corresponding to this silhouette---the \visibility\ boundary is the set of points at which the surface normal is orthogonal to the view ray to that point. As in the intersection optimization stage, we attempt to optimizate all points with respect to this objective, and label those which converge to within 5 pixels of their starting position as \visibility\ boundaries, updating their positions according to the optimization result. Note that we retain any and all intersection objectives in this step, as a vertex may lie at both the intersection of surfaces and the \visibility\ boundary of a surface (see the two purple points in Figure \ref{fig:wireframe} (e)).

The optimization steps displaces vertices, causing some artifacts in the wireframe, so we perform final wireframe post-processing and surface mesh extraction to generate the final clean VB-Rep; see Sections \ref*{sec:postprocess} and \ref*{sec:mesh} of the supplemental material for details.
\section{Experimental Results}

We evaluate our method on RGB-D images of man-made objects of varying complexity from the synthetic dataset described in Section \ref{sec:methods}, as well as some real-world RGB-D measurements obtained with an Intel® RealSense™ depth camera.

\subsection{Qualitative assessment}
\begin{figure*}
    \centering
    \includegraphics[width=\textwidth]{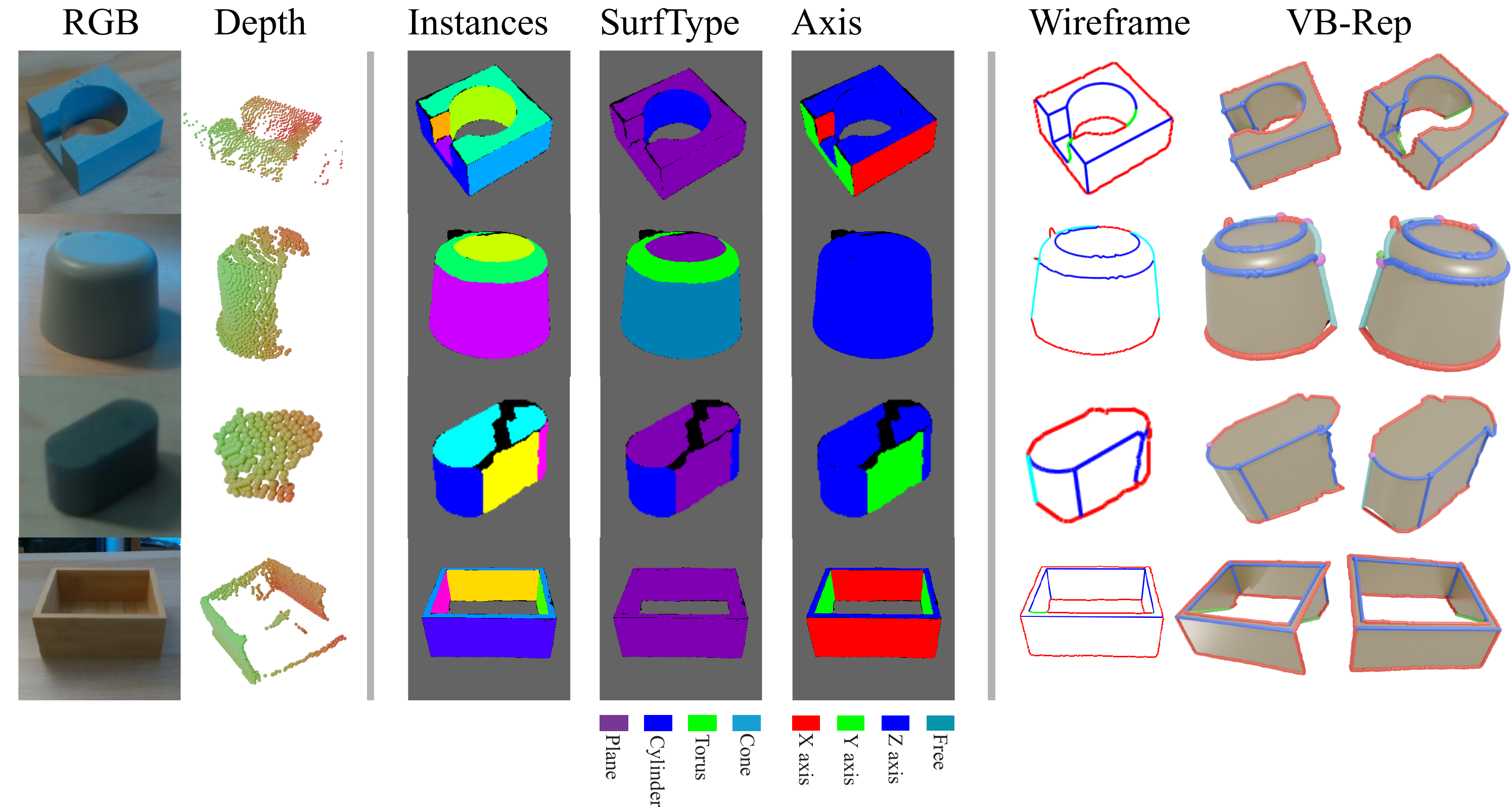}
    \caption{Results of our segmentation and VB-Rep extraction on various real objects. Real RGB-D inputs were captured using a RealSense™ camera. All edges in the resulting VB-Reps are colored according to their type: \intersection\ edges are blue, \silhouette\ edges are red, \visibility\ edges are cyan, \occluded\ edges are green; vertices are colored the same way. In addition, triple intersection vertices and \intersection-\visibility\ vertices are shown as balls, with the latter colored magenta.}
    \label{fig:realresults}
\end{figure*}

\vspace{2pt}\noindent\textbf{Experiments with real data}
Results of our method on real-world RGB-D captures are shown in Figure~\ref{fig:realresults}. Due to the fact that the RGB and depth cameras do not share the same viewpoint, we must take additional steps to reproject the depth point clouds to the viewpoint of the RGB camera so that we can associate the correct subsets of points in $\pointcloud$ with the predicted segmentation regions; this introduces some artifacts, such as holes in the depth points at occlusion boundaries. Considerable artifacts and holes can also be seen in the depth measurements themselves. 

The combination of intersection and alignment constraints makes the method quite robust; despite extreme errors and lack of detail in the depth point cloud, the reconstructed models exhibit clean topology (up to the limits of visibility). 
While some errors exist in the segmentation due to the sim-to-real gap between our training data and the captured objects, our algorithm is robust enough to correct many of them (such as the split plane in the third row), while minimizing others (the jutting boundary in the second row). In other cases, the method fails ``gracefully'' (the missing floor of the box in the fourth row), allowing good quality reconstruction of all detected geometry.

\begin{figure*}
    \centering
    \includegraphics[width=\textwidth]{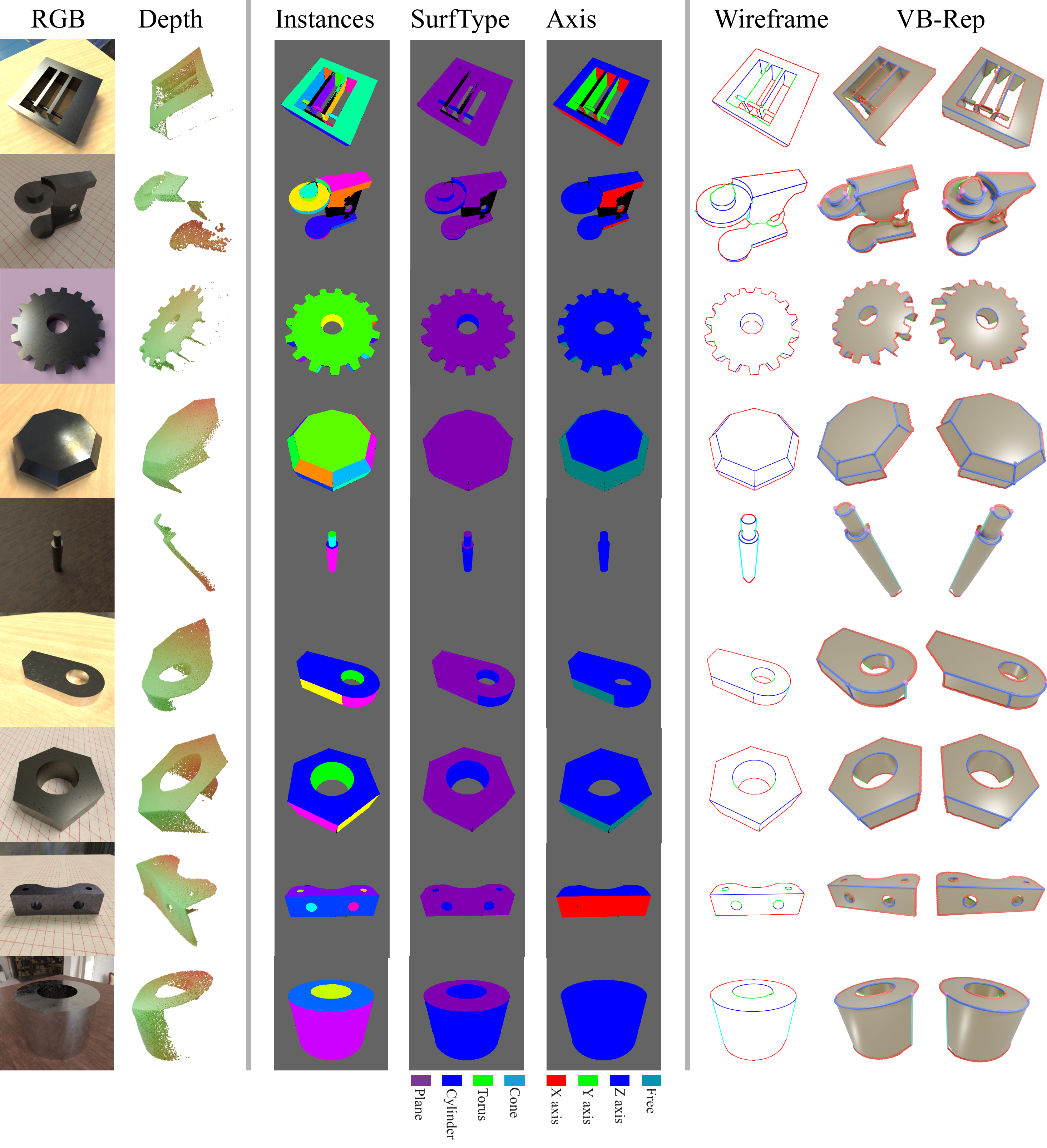}
    \caption{Results of our segmentation and VB-Rep extraction on additional synthetic examples. Synthetic depth maps are perturbed using FBM noise to simulate sensor errors.}
    \label{fig:syntheticresults}
\end{figure*}

\vspace{2pt}\noindent\textbf{Synthetic Validation}
\label{sec:synthetic}
We further validate our method using the synthetic dataset. To simulate sensor errors, we use a combination of fractal brownian motion (FBM) noise and bilateral gaussian blur for a similar effect to the systematic errors and over-smoothing we observe with the RealSense™ camera. 
Figure \ref{fig:syntheticresults} shows the results of our method on several synthetic inputs. 
On some of the more complex models (the fifth and sixth rows) even when the segmentation model misses certain regions, plausible geometry is inferred outside of these localized errors.

The blue \intersection\ edges and vertices in Figures \ref{fig:realresults} and \ref{fig:syntheticresults} show which VB-Rep elements can be identified as part of the original CAD B-Rep geometry. Since they are constrained by neighboring optimized surface geometry, they form clean, precise curves. Other types of edges show the various ways parts of a model can be hidden from view.

\subsection{Quantitative evaluation}
\label{sec:metrics}



\begin{figure*}
    \centering
    \includegraphics[width=0.5\textwidth]{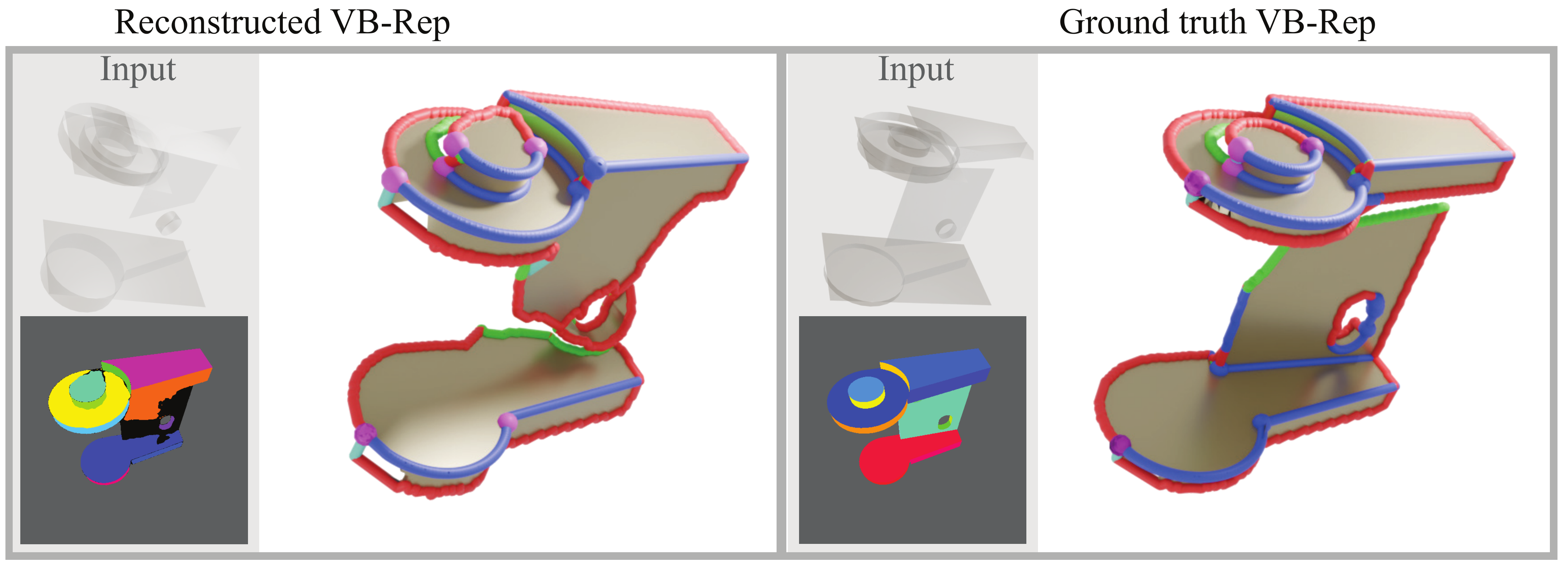}
    \caption{Reconstructed and ground truth VB-Rep. For the ground truth VB-Rep, we use the known set of surface primitives and corresponding pixel instance map as input to the wireframe/mesh extraction steps. }
    \label{fig:gtvbrep}
\end{figure*}

To measure the geometric reconstruction accuracy on a synthetic validation set of 41 examples, we use the \textbf{Chamfer distance (CD)} between the reconstructed VB-Rep geometry and the ground truth VB-Rep and \textbf{Primitive alignment} for individual primitives, measured as Euclidean distances between primitive axis. 
Primitive alignment is between corresponding pairs of primitives, so we compute a greedy matching across both sets of primitive instances according to their overlapping IoU in image space, stopping at 75\% IoU. Averaged results are shown in Table \ref{tab:results}. The ground truth VB-Reps used for these metrics are produced by running just the wireframe and mesh extraction from Section~\ref{sec:wireframe} starting with the ground truth primitives from $\truebrep$ (see Figure~\ref{fig:gtvbrep}).

We also compute a \textbf{face matching precision/face matching recall} based on this matching, which measure the proportion of ground truth faces that are matched by a predicted face, and the proportion of predicted faces that are matched by a ground truth face, respectively. Given the ground truth geometry of our synthetic examples, we evaluate the performance of primitive instance detection using standard classification metrics over the set of primitives: \textbf{mean primitive type accuracy} and \textbf{axis type accuracy}. We find the average values of these \textbf{segmentation-focused} metrics are as follows: face matching precision: 88.1\%; face matching recall: 91.3\%; axis accuracy: 94.5\%; type accuracy: 99.2\%. 

The runtime of our method is about 50 seconds per example on average.



\begin{figure}
    \centering
    \includegraphics[width=0.5\textwidth]{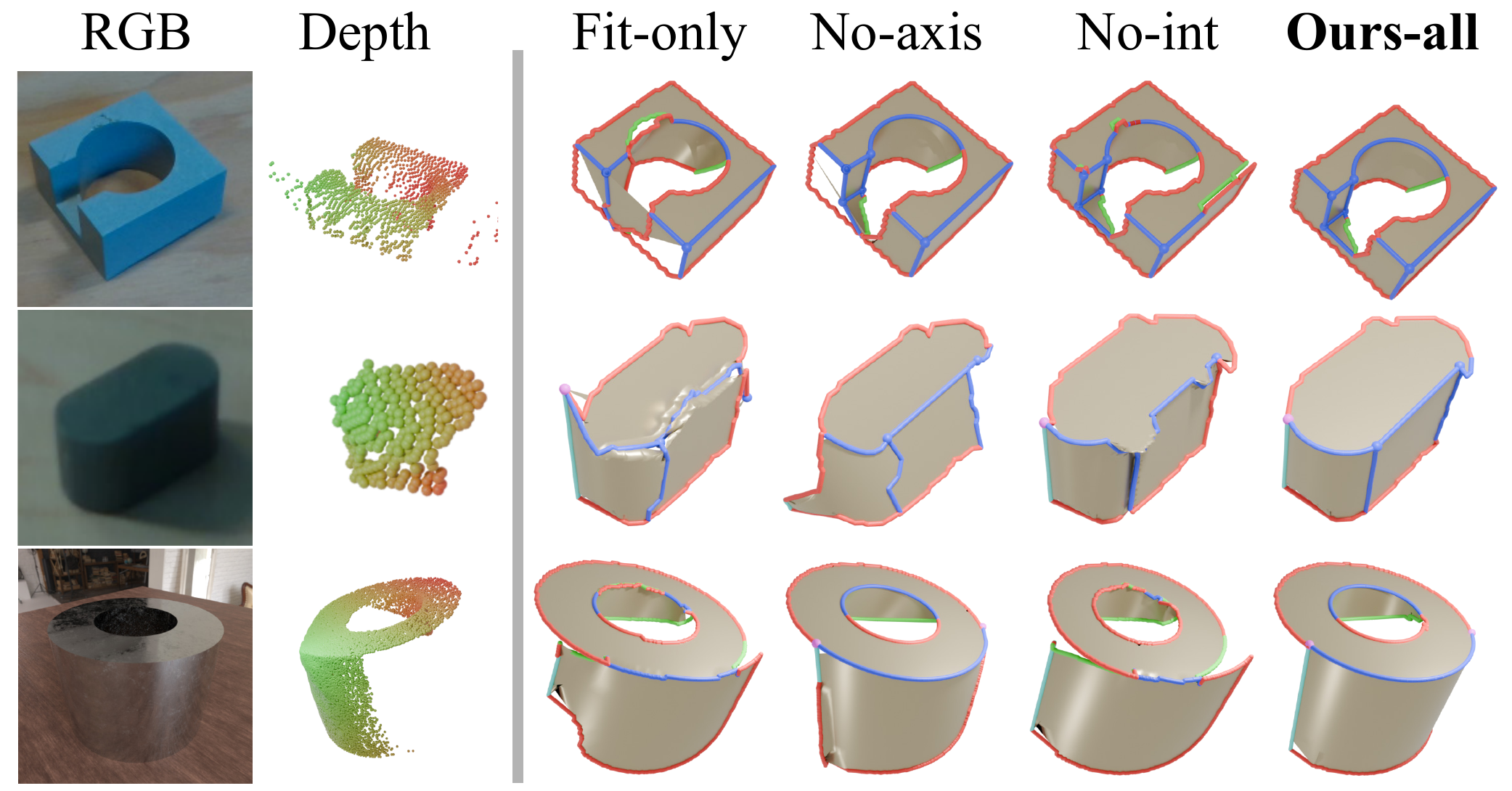}
    \caption{Comparison of results with our full pipeline with individual features left out, on real data (first two rows) and a synthetic example (third row) with simulated depth error.
    }
    \label{fig:ablations}
\end{figure}

\vspace{2pt}\noindent\textbf{Ablation study}
\label{sec:ablation}
We examine the effect of holding out various parts of our algorithm on the quality of results in Figure~\ref{fig:ablations}. With \textbf{No-int}, we leave out $E_{\mathrm{int}}$ from the optimization; with \textbf{No-axis}, we do not constrain axes based on predicted alignment during primitive fitting; and \textbf{Fit-only} uses neither of these refinements. Disabling intersection guidance leads to many tears in the model, showing that the intersection term is crucial for ensuring continuity of the VB-Rep. However, the intersection term alone cannot ensure a consistent CAD model, as without axis guidance, features features become misaligned, sometimes causing broken topology as in the top example. 

Quantitative effects of the above modifications (with the synthetic errors described in Section \ref{sec:metrics}) are shown in Table~\ref{tab:results}. We observe that \textit{ours-all} outperforms all other ablations, and that axis guidance is essential for predicting correct primitive orientations. 


\begin{table}
    \caption{Ablation study: We evaluate our method with various features removed as described in Section \ref{sec:ablation} and average the results over 41 examples. All experiments use synthetic sensor errors as described in Section \ref{sec:synthetic}.}
    \centering
    \begin{tabular}{c c c }
    \toprule
       Method   &  CD($\downarrow$)  & Primitive alignment($\downarrow$)   \\
    \midrule
        \textbf{Ours-all} & \textbf{0.0628} & \textbf{0.0693}  \\
        No-axis & 0.101 & 0.251   \\
        No-int & 0.0768 & 0.0940  \\
        Fit-only & 0.0819 & 0.244 	   \\

    \end{tabular}
    \label{tab:results}
\end{table}

\subsection{Comparisons with Prior Work}
\begin{figure}
    \centering
    \includegraphics[width=0.45\linewidth]{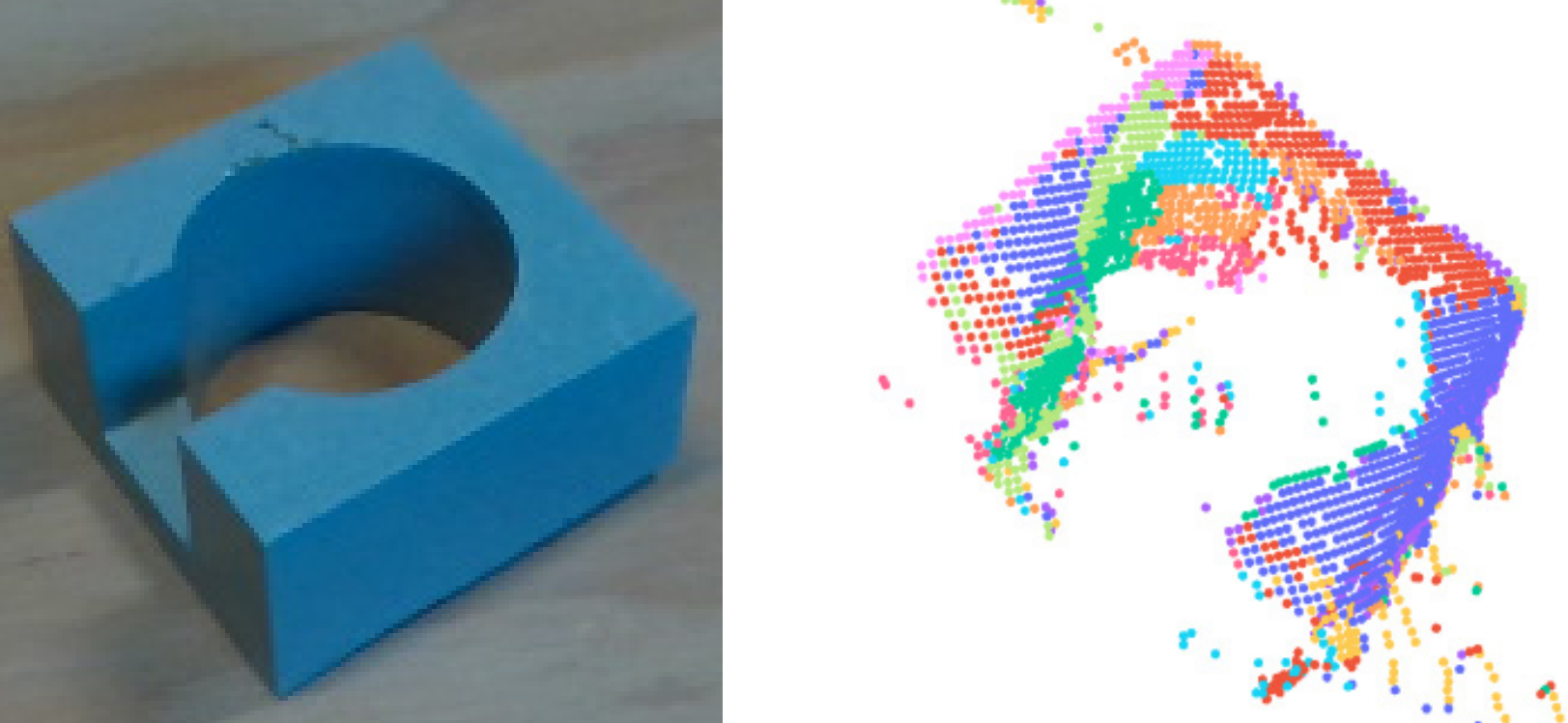}
    \caption{3D scan segmented using HPNet\cite{yan2021hpnet}.}
    \label{fig:hpnet}
\end{figure}

Since there is no directly applicable method for the problem of view-centric CAD reconstruction from a single RGB-D image, we examine prior works for reconstructing B-Reps from point clouds. We compare with Point2CAD~\cite{liu2024point2cad} by running the primitive segmentation backbone model HPNet~\cite{yan2021hpnet} on the point could generate by our depth sensor, with results shown in Figure \ref{fig:hpnet}. Even though we removed background points and normalized the remaining points to the centered unit cube in an attempt to match the expected input, the result is heavily over-segmented and B-Rep extraction cannot proceed.

We also applied a recent work, Split-and-Fit~\cite{liu_sig24}, but it could not produce any output on our data, as the model is trained only on clean, synthetic point clouds.

\section{Discussion and Future Work}

\begin{wrapfigure}{r}{120pt}
    \centering
    \vspace{-10pt}
    \hspace*{-20pt}\includegraphics[width=140pt]{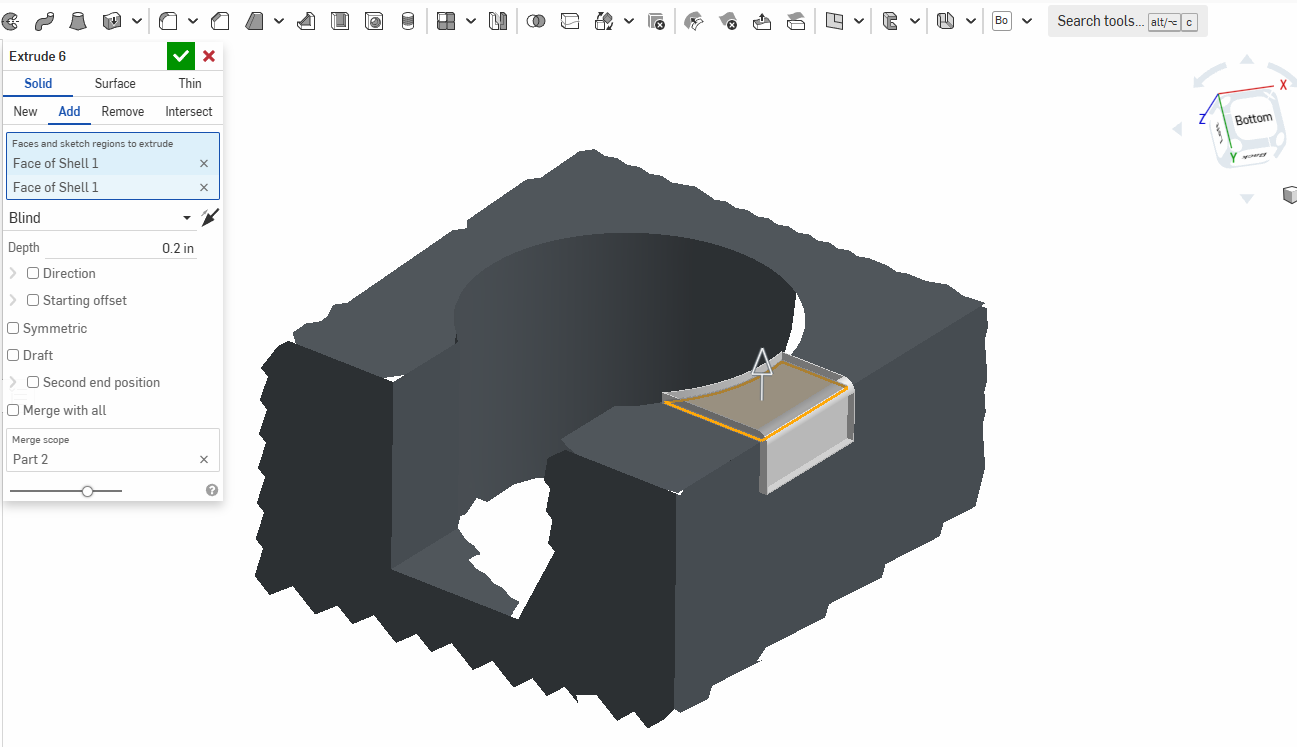}
    \vspace{-10pt}
    \label{fig:illustrative}
\end{wrapfigure}In this paper, we present a novel method for partially reconstructing CAD shapes from single RGB-D images, effectively addressing the challenges posed by incomplete and ambiguous topological information through the incorporation of visibility limits and geometric uncertainty. This approach can enable broad practical applications, including reconstructing accurate models for analyzing manufactured objects, designing fixtures, attachments, or custom components that seamlessly integrate with existing objects (see inline figure), and guiding the creation of manufacturable designs with variations derived from the original.

Future work could extend this method to more complex surfaces, such as B-splines, and introduce additional priors, like tangency constraints, to enhance accuracy. While our segmentation approach can generalize to real data, further modeling of real-world phenomena could improve resilience in diverse settings. Additionally, eliminating the need for depth input by building on single-view depth estimation works could extend the method’s applicability. 

Further work should also focus on integrating this approach seamlessly into user workflows, such as by developing plugins for CAD systems that can directly interface with VB-Reps to take advantage of the extra uncertainty information that our representation provides.

Another exciting area for future work involves incorporating multiple input images, which are often available in practical applications. Even with multiple views, however, occlusions remain inevitable in most scenarios. Our work, as the first to address the challenge of extracting B-Reps with incomplete model topology while leveraging RGB information rather than relying solely on point clouds, provides a foundation for future methods that aggregate multiple views while effectively handling partial observations.

\clearpage
{
    \small
    \bibliographystyle{ACM-Reference-Format}
    \bibliography{bib}
}

\clearpage
\pagenumbering{arabic} 
\setcounter{page}{1} 
\setcounter{figure}{0} 
\setcounter{section}{0}

\section*{Supplemental Material} 

\section{VB-Rep Implementation}
As with regular B-reps, each element in the VB-rep graph is associated with geometry. For faces, we use parametric surfaces as in classic B-reps as these can be fitted by a reconstruction algorithm. In this work, we use five of the standard primitives: \textbf{planes}, \textbf{cylinders}, \textbf{spheres}, \textbf{cones}, and \textbf{tori}. Intersection edges are also associated with parametric curves defined by the intersection of two surfaces. The notion of an intersection curve is a representation supported by common B-Rep packages, such as Parasolid~\cite{parasolid}. For the other types of curves, there is no analytic geometric representation like the intersection of two surfaces, so we fall back on our polyline approximation. Finally, vertices are represented as standard points.

In our implementation, we approximate all edges using densely-sampled piecewise line segments (due to the difficulty of representing geometries with only one constraining surface), and classify our boundaries at the level of these segments. 

The vertices in our VB-Rep possess the same classifications, with the caveat that a vertex may be both an \intersection\ vertex and a \visibility\ vertex (see the magenta point in Figure \ref*{fig:vertices} of the main paper), and vertices with three intersecting surfaces (blue circle in the figure) are \textbf{corners} which correspond with vertices in $\truebrep$.

\section{Dataset\label{sec:dataset}}
\label{sup:dataset}
Our CAD primitive segmentation dataset consists of synthetic renders of 50,000 single-part CAD models from the AutoMate Part Dataset~\cite{automate}, consisting of plane, cylinder, sphere, cone, and torus primitives. In order to take steps towards closing the sim-to-real training gap, we render our images with randomized lighting, material, and camera viewpoint using the physically-based path tracer Mitsuba~\cite{mitsuba}, and along with each image, we render ground truth labels associating pixels to primitive instances. The dataset includes the labeling of each of the instances into one of the five primitive types; in addition, we include labels for each primitive's primary orientation axis.

\paragraph{Scene setup}
Each of our scenes is comprised of a single ground plane with random rectangular bounds with the CAD part placed in the center, and the camera looking inwards from a random point sampled on a sphere of radius $r_\mathrm{view}$. For lighting, we use a set of 10 randomly chosen environment maps downloaded from Poly Haven, and for materials, we use a collection of 19 PBR materials for different kinds of wood and metals downloaded from \url{3DTextures.me} complete with color, normal, roughness, and metallic maps. We also randomize the colors for a subset of these materials to provide some additional variation to the training data. We choose a natural ``resting'' orientation for the object based on maximal contact with the ground plane, in order for our scenes to resemble a plausible real capture scenario. We render the scenes with Mitsuba using the standard \textbf{path} integrator for simulating realistic multiple-bounce lighting. 

\paragraph{Axis labels}
Each primitive can be labeled as aligned with the X axis, Y axis, Z axis, or no axis if its orientation is unconstrained. Spheres have no orientation, so multiplying each of the remaining primitive types by these orientation choices gives us a total of 17 types to classify. Since all of the models are scraped directly from CAD modeling software, axis aligned features are almost always aligned solely with the model's local coordinate axes, allowing us to easily identify the alignment of each primitive using a small threshold (we use $1.5^\circ$ angle difference between the primitive axis and the coordinate axis).  However, these labels pertain to the \textit{model} coordinate frame, which is arbitrary since for single-view reconstruction, rotating the model is no different from changing one's viewpoint. We therefore redefine the X axis to be whichever of the model coordinate axes is closest to parallel with the view (so that features facing the view head-on are aligned with X), and leave Z pointing up to obtain the final, view-corrected orientation labels.

\section{VB-Rep Post-Processing}
\label{sec:postprocess}

In the process of VB-rep extraction, we have classified parts of our wireframe as \intersection, \visibility, \silhouette, and \occluded\ at the level of vertices, rather the segments joining them. To obtain a VB-Rep with edge structures labeled according to the various visibility phenomena, we must precisely label the segments in our wireframe.  

Furthermore, the optimization steps of our surface-aware wireframe extraction method, especially the visibility optimization, may displace the vertices of the wireframe considerably, leading to artifacts that must be fixed via post-processing. 

We detail these two steps below. 

\paragraph{Edge labeling}  We adopt the following rules: An edge segment is an \textbf{\intersection} edge if both neighboring vertices are \intersection\ vertices; an edge segment is a \textbf{\visibility} edge if both neighboring vertices are \visibility\ vertices; an edge segment is \textbf{\occluded} if at least one of its vertices is not an \intersection\ vertex, and at least one of its vertices is an \occluded\ vertex. If our piecewise linear approximation of the VB-Rep edges is taken to the limit where segments are infinitesimal and all surfaces locally planar, these rules hold exactly.

\begin{figure}
    \centering
    \includegraphics[width=\linewidth]{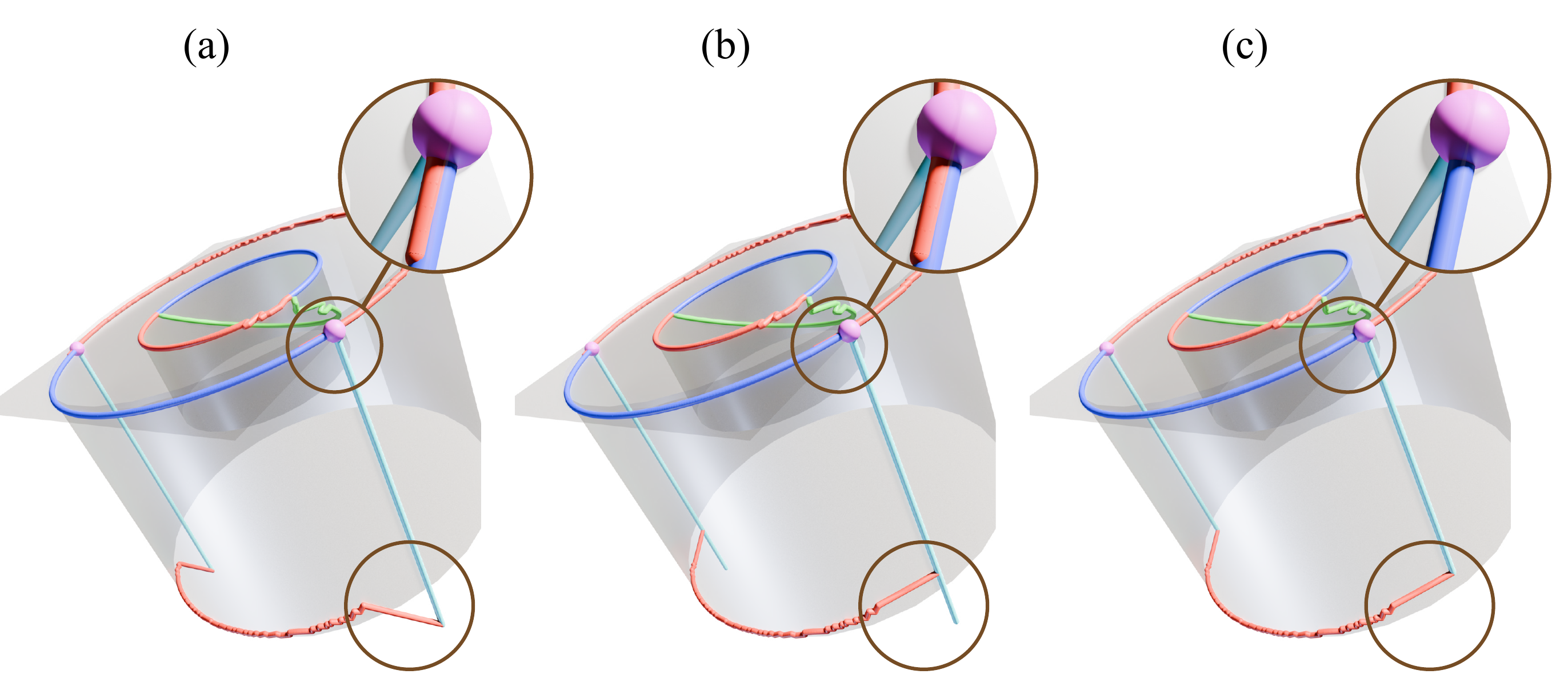}
    \caption{Wireframe post-processing steps. Visibility optimization displaces vertices, leading to protruding corners at the endpoints of visibility edges and overlapping edges at visibility-intersection corners (a). We can minimize the displacement at the endpoints of visibility edges (b), followed by removing degenerate vertices (c) to obtain a clean wireframe without artifacts.}
    \label{fig:postprocessing}
\end{figure}

\paragraph{Wireframe post-processing} The optimization steps described in Section~\ref*{sec:wireframe} may lead to some artifacts, since they act on the wireframe vertices without regard for their connectivity, as shown in Figure~\ref{fig:postprocessing}. In particular, since the depth gradients near visibility boundaries are high, the visibility optimization step may displace vertices considerably along the viewing direction, leading to protruding corners as in the bottom right of part (a) of the figure. We fix these artifacts in two steps: \textbf{corner correction} and \textbf{wireframe simplification}. Corner correction seeks to minimize the displacement of the ``unconstrained'' endpoints of \visibility\ edges (bordering \silhouette\ edges), shown in the lower right of Figure~\ref{fig:postprocessing} (a). To do this, we find all vertices adjacent to both a \visibility\ and a \silhouette\ edge and, for each such vertex, move the vertex to the orthogonal projection of the other endpoint of the \silhouette\ edge onto the visibility edge. This leads to some overlapping edges, however (see Figure~\ref{fig:postprocessing} (b)), so we perform a wireframe simplification step which fixes both these and other overlaps caused by the intersection-visibility vertices, with the result shown in Figure~\ref{fig:postprocessing} (c). Specifically, we post-process the wireframe by deleting vertices with degenerate edge angles (in practice, less than $36^\circ$) iteratively until none are left; we find that this fixes most geometric artifacts arising from the above steps. 


\section{Surface Mesh Extraction}\label{sec:mesh}
\begin{figure}
    \centering
    \includegraphics[width=0.9\linewidth]{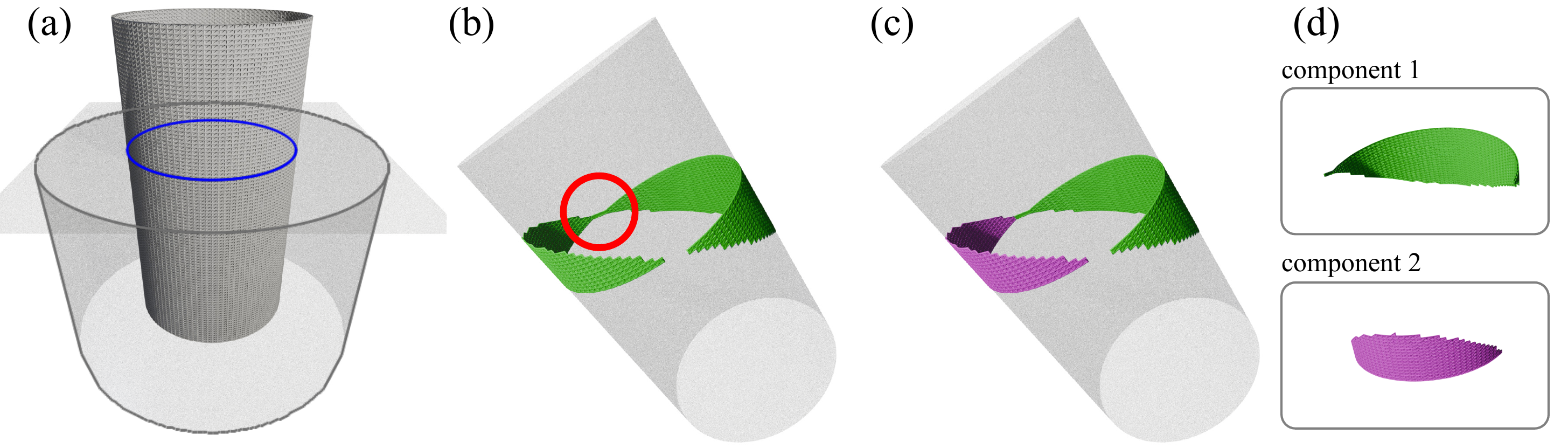}
    \caption{Visibility mesh extraction. Given the 2D boundary of a particular surface $\surface$ (the blue curve in (a)), we extract the submesh of $\surface$ with vertices whose projection is contained in the 2D boundary (b). Mutually exclusive regions may be incorrectly joined in this mesh (see the red circle), so we label the submesh faces according to their view normal (c), and use connected components on the face adjacency graph and the orientation labeling to extract all bounded submeshes (d).}
    \label{fig:meshes}
\end{figure}

For extracting the visible surfaces $\visiblemesh$ in Section \ref*{sec:wireframe} as well as our final VB-Rep surfaces, we use a discrete mesh processing algorithm which leverages the assumption that all surfaces in the VB-Rep are bounded by their visual extent from a particular viewpoint. Our procedure is illustrated in Figure \ref{fig:meshes}: First, we generate a mesh $\hat\visiblemesh_i$ for $\surface_i\in\visiblesurfaces$ by mapping a high-resolution square grid mesh to 3D using the surface parameterization from UV space to 3D $F_i:\mathbb{R}^2\to\mathbb{R}^3$. We extract the vertices of $\hat\visiblemesh_i$ that are contained in the 2D boundary of the region corresponding with $\surface_i$, illustrated by the blue curve in \ref{fig:meshes} (a), resulting in a submesh $\bar\visiblemesh_i$ shown in Figure\ref{fig:meshes} (b). To ensure that the distinct depth layers of $\surface_i$ are correctly separated, we also label the faces by their normal dot product with the view ray (Figure \ref{fig:meshes} (c) and use a connected components algorithm on the face adjacency graph of $\bar\visiblemesh_i$ with this face labeling to extract our final visibility meshes $\visiblemesh_i^k$ for $k=1\dots N_{\mathrm{depths}}$.

Note that the surfaces $\visiblesurfaces$ may be unbounded. When generating $\hat\visiblemesh_i$ using a finite square grid in UV space, we adaptively scale the relevant dimensions (height for cylinders and cones, width and height for planes) until the contained mesh $\bar\visiblemesh_i$ contains no edges from the UV grid boundary (indicating where our mesh cuts off prematurely). 

For visualization purposes in the final surface meshes, we close the gap between the mesh edges and the boundary wireframe by projecting the non-manifold edges of this mesh to the nearest point on the wireframe.

\section{Parallel Plane Rejection}
\begin{figure}
    \centering
    \includegraphics[width=0.9\linewidth]{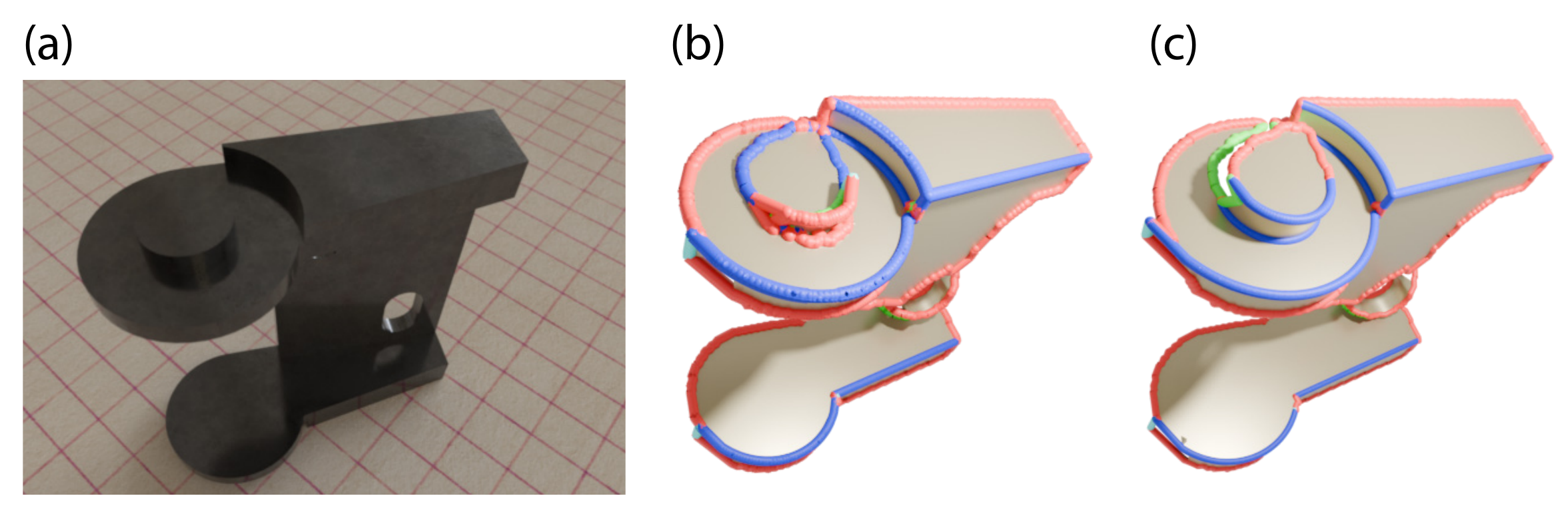}
    \caption{With an input object (a) containing slightly offset plane features, intersection refinement can lead to an incorrect result (b); note how the top stud gets ``glued'' to the plane below. We can explicitly use the assumption that parallel planes cannot intersect (c), fixing this common failure case.}
    \label{fig:planereject}
\end{figure}

Due to the use of a fixed distance threshold $\intthresh$ during the intersection discovery process described in Section~\ref*{sec:geomopt}, it can happen that surfaces that are merely close, as well as adjacent to a shared edge in the 2D edge graph, are erroneously classified as intersections, leading the refinement to produce an incorrect result (see Figure \ref{fig:planereject} (b)). We find that a common case where this happens is in models with slightly extruded surfaces, where a plane borders an offset plane in the 2D image, and the offset between them is smaller than $\intthresh$. To avoid this common failure case, we explicitly use the knowledge that parallel planes do not intersect, and since we predict whether any two planes should share an axis in the previous steps, we can exclude any vertices neighboring two planes with the same predicted axis (provided that the predicted axis is not unconstrained) from consideration in the intersection refinement step. The resulting improvement is shown in Figure \ref{fig:planereject} (c). 

\section{Synthetic Noise Model}
\begin{figure}
    \centering
    \includegraphics[width=0.5\linewidth]{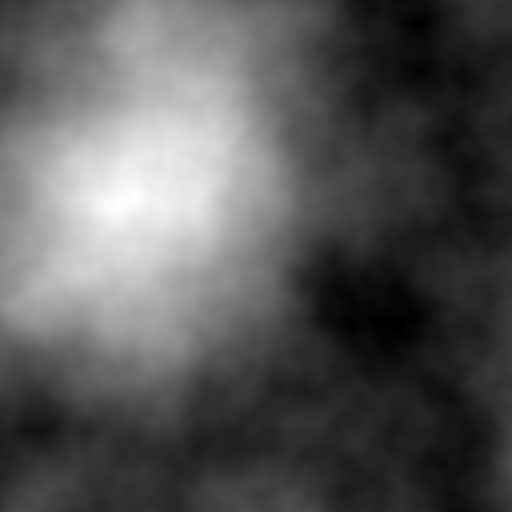}
    \caption{Fractal Brownian Motion noise used to perturb the synthetic depth maps.}
    \label{fig:fbm}
\end{figure}

The noise model we use to corrupt our synthetic data for the purpose of our quantitative ablation study in Section~\ref*{sec:ablation} is \textbf{Fractal Brownian Motion} (Figure~\ref{fig:fbm}). We use a standard implementation based on Perlin Noise, using the parameters $\mathrm{lacunarity}=2.0$, $\mathrm{persistence}=0.5$, $\mathrm{octaves}=7$, $\mathrm{scale}=1000$. We map this noise to the range $[-0.25, 0.25]$ and add it to the synthetic depth values. In our ablations, we use an identically-seeded noise map to fairly compare the different methods.

\section{Axis Reparameterization for Optimization}
In Section~\ref*{sec:geomopt}, we jointly optimize the primitive parameters and the axis-angle rotation frame $\mathcal{R}$. $\mathcal{R}$, as well as many of the surface primitive parameters, contains unit vectors that must remain unit length throughout the optimization (such as the axes of cylinders, cones, and tori). Rather than treat these as 3-component vectors for purpose of optimization, which would require re-normalization steps during optimization, we reparameterize all unit vectors in terms of two variables. Given a unit vector $\hat{\mathbf{a}}$, we define a perturbed vector 
\begin{equation}
\mathbf{a}(u, v) = \hat{\mathbf{a}} + u\cdot\hat{\mathbf{n}} + u\cdot\hat{\mathbf{b}}
\end{equation}
where $\hat{\mathbf{n}}$ and $\hat{\mathbf{b}}$ are perpendicular unit vectors, spanning the plane orthogonal to $\hat{\mathbf{a}}$. Finally, we use the value of $\mathbf{a}/\|\mathbf{a}\|$ in place of $\hat{\mathbf{a}}$ wherever we have unit vectors in our optimization.





\end{document}